\documentclass[aps,prl,twocolumn,showpacs,amsmath,amssymb]{revtex4}

\usepackage{graphicx}
\usepackage{dcolumn}
\usepackage{bm}

\begin{document}

\title{Molecular Probe of Pairing in the BEC-BCS Crossover}

\author{G. B. Partridge}
\author{K. E. Strecker}
\altaffiliation{Sandia National Laboratories, Livermore, CA 94551}
\author{R. I. Kamar}
\author{M. W. Jack}
\author{R. G. Hulet}
\affiliation{Department of Physics and Astronomy and Rice Quantum
Institute, Rice University, Houston, Texas 77251}

\date{\today}

\begin{abstract}
We have used optical molecular spectroscopy to probe the many-body
state of paired $^6$Li atoms near a broad Feshbach resonance. The
optical probe projects pairs of atoms onto a vibrational level of an
excited molecule.  The rate of excitation enables a precise
measurement of the closed-channel contribution to the paired state.
This contribution is found to be quite small, supporting the concept
of universality for the description of broad Feshbach resonances.
The dynamics of the excitation provide clear evidence for pairing
across the BEC-BCS crossover, and into the weakly interacting BCS
regime.


\end{abstract}


\maketitle

Feshbach resonances were recently used to explore the
possibility of superfluidity and Cooper pairing in atomic Fermi
gases of $^{40}$K and $^6$Li. Two experiments have produced evidence
for correlated pairs \cite{Regal04AZwierlein04}, while
 others report evidence for superfluidity \cite{Kinast04Bartenstein04aKinast05}. These experiments are noteworthy because
they are performed on the high-field side of the resonance, where
the $s$-wave scattering length $a$ is negative and, according to
two-body physics, diatomic molecules associated with the resonance
are energetically unstable. These intriguing experiments raise
important questions regarding the nature of Feshbach-induced pairs,
including their relation to bound molecules and the role played by
many-body effects.

Atomic Feshbach resonances involve a collision of a pair of atoms in
an open channel coupled to a bound molecular state of a closed
channel \cite{duine04}. In the broad $^6$Li resonance, the open channel
corresponds to two atoms interacting mainly via the electronic
spin-triplet potential, while the closed-channel is predominantly
spin-singlet. Two-body theory predicts the formation of a weakly-bound
molecular state on the ``BEC" side resonance, where $a>0$. These
``dressed" molecules are superpositions of open-channel atoms with the
``bare" molecules of the closed channel. For resonances that are broad
compared to the Fermi energy, the closed channel character of the
dressed molecules is expected to be small throughout an experimentally
relevant region about the resonance
\cite{Bruun04depalo04Diener04Simon04}. In this case, the resonance may
be well-described by a single-channel model where the physics is
universal, such that the macroscopic properties of the superfluid are
independent of the microscopic physics that underlie the two-body
interactions. Universality in the description of pairing in atomic gases
establishes their relevance to other systems, most notably
high-temperature superconductors. 

In the experiment reported here, a laser is used to project the
dressed molecules/pairs onto an excited singlet molecular state. By
starting with an evaporatively cooled gas on the BEC side of the
resonance, followed by an adiabatic change in the magnetic field, a
nearly zero temperature gas can be probed throughout the BEC-BCS
crossover. This enables a direct measurement of the closed
channel fraction and,  for the first time in an atomic gas,  provides clear
evidence for the presence of pair correlations in the weakly interacting
BCS regime.

Our apparatus and methods for producing strongly interacting Fermi
gases of $^6$Li atoms have been described previously
\cite{streck03}. The atoms are confined in a single-beam optical trap which, at full laser intensity, has a trap depth of 25 $\mu$K and radial and axial freqencies of  $\nu_r = 2270$ Hz and $\nu_z = 21$ Hz, respectively. Curvature in the
magnetic bias field modifies the axial frequency such that
$\nu_z' = \sqrt{\nu_z^{2}+\lambda B}$,
where $\lambda = .029(3)$ Hz$^2$/G. An incoherent equal mixture of
the two lowest going Zeeman sublevels ($F = 1/2, \, m_F = \pm 1/2$)
is created \cite{streck03}. For the magnetic fields of interest
($B \geq 600$ G) these states are nearly electronically spin
polarized, but differ in their nuclear spin projections. They
interact strongly via a broad Feshbach resonance located near 834 G
\cite{Houb98,Bartenstein05}.

\begin{figure}
 \includegraphics[scale=0.35, bb=50 179 712 729]{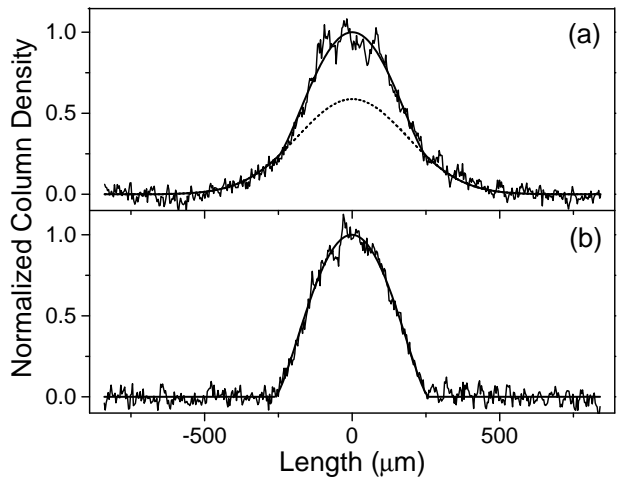}
\caption{\label{mbec} In-situ absorption image profiles showing a
molecular BEC.  These images were recorded at a field of 695 G after
evaporation at 754 G.  For (a), the optical trap depth was lowered
to 0.5 $\mu$K. The solid line is a fit to a Gaussian
(dotted line) plus Thomas-Fermi distribution which distinguishes the
condensate from residual thermal molecules. For (b), the trap depth is reduced to 0.27
$\mu$K, producing an essentially pure molecular condensate with the
number of molecules, $N = 46,000$. We estimate that the condensate
fraction is $>$$90\%$, implying $T/T_c \leq 0.5$, where $T_c$ is the
critical temperature for BEC at 695 G. The solid line is a
Thomas-Fermi distribution.}
\end{figure}

A molecular BEC is created  as in other $^6$Li
experiments \cite{JochimSci03Zwie03Bourdel04}.  We prepare the spin
mixture at 754 G (corresponding to $a \simeq 3680~ a_o$
\cite{footnote_a_o}), and dressed molecules form through three-body
recombination. The molecules are evaporatively cooled by reducing
the trap depth in 750 ms in an approximately exponential trajectory.
Their distribution is imaged in the trap with the same optical probe
used to image the atoms. Figure \ref{mbec} shows that the condensate
fraction can be controlled by the final trap depth. The crossover
regime is probed by preparing the gas at 754 G and adiabatically
changing the magnetic field to any desired final value.  We find
that adiabatic field excursions, even those going across the
resonance, result in no detectable heating upon returning to the
original field \cite{Bartenstein04b}.

A molecular probe laser drives transitions between the dressed
molecules/pairs and an electronically excited molecular singlet
level.  These transitions result in spontaneous emission and a
detectable loss of atoms from the trap. We chose to drive
transitions to the $v' = 68$ level of the $A^1 \Sigma _u ^+$ excited
state because it has the largest Franck-Condon wavefunction overlap
(0.077) with the resonant channel bare molecule, the $X^1 \Sigma _g
^+(v=38)$ level. The $v' = 68$ level has a classical turning point of 36 $a_{0}$. A quantity $Z$ is defined in terms of $\Gamma$,
the rate of photoexcitation, by $\Gamma = Z
\Omega^2/\gamma$, where $\Omega = \langle \psi_{v'=\,68}(S=0)|
\vec{d} \cdot \vec{E_{\rm{L}}} | \psi_{v=\,38}(S=0) \rangle $ is the
on-resonance Rabi frequency, $\vec{d}$ is the transition dipole,
$\vec{E_{\rm{L}}}$ is the laser field of the molecular probe, and
$\gamma = (2\pi)~11.7$ MHz is the linewidth of the excited molecular
state \cite{Prodan03}. The dressed
molecules/pairs can be expressed as a superposition of the $v=38$ singlet
molecules and free atom pairs in the triplet channel \cite{duine04}:
\begin{equation}
|\psi_{\rm p}\rangle=Z^{1/2}|\psi_{v=\,38}(S=0)\rangle +
(1-Z)^{1/2}|\phi_{\rm a}(S=1)\rangle, \label{eq:dressed_mol}
\end{equation} and $Z$ can be identified as the component fraction of the dressed
molecules in the closed-channel. For a typical probe laser intensity
of 20 mW/cm$^2$, $\Omega = (2\pi)~2.6$ MHz \cite{Prodan03}. 

$Z$ is measured at various fields by the
following process: first, the molecular gas is prepared at 754 G.
The field is then linearly ramped to a new value $B$ in 160 ms and
held for 20 ms before the molecular probe is pulsed on for a fixed
duration. The probe removes atoms at a rate
$\Gamma$ determined by $Z$. The field is then linearly returned to
754 G in 160 ms, where the number of remaining atoms is determined
by optical absorption imaging. The measured numbers are normalized
to data obtained by the same procedure, but without firing the
molecular probe.

\begin{figure}
 \includegraphics[scale=0.35, bb=50 179 712 729]{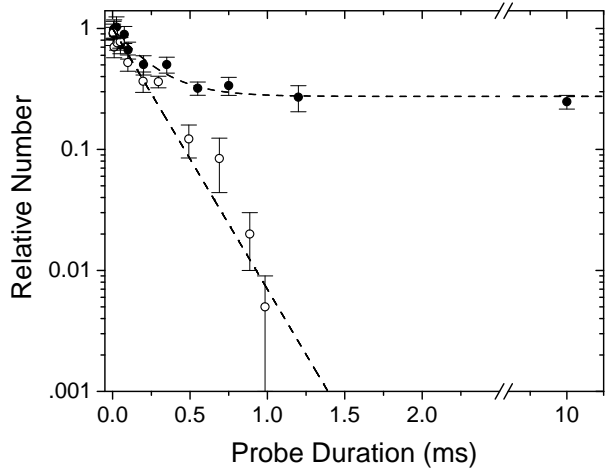}
\caption{\label{loss695} Loss of signal vs.\! molecular probe
duration at 695 G. The open circles correspond to a gas
evaporatively cooled to a nearly pure molecular BEC, while the
closed circles correspond to full trap depth, where $T/T_F \simeq
0.75$. The dashed lines are fits to exponentials, with a leftover
fraction of 25\% in the  high temperature case.   The time
axis for the BEC data was scaled to account for differences in
molecular probe laser intensity between the BEC data (30 mW/cm$^2$)
and the high temperature data (15 mW/cm$^2$). We have verified that
the loss rate depends linearly on intensity.  The error bars
represent the statistical standard deviation from the mean of
$\sim$10 independent measurements. }
\end{figure}

Figure \ref{loss695} shows the normalized loss of atoms at $B = 695$
G as a function of the molecular probe duration for two different
temperatures.  In the case of the pure condensate, the gas is well
into the BEC regime, since $a \simeq 1510 ~a_o$ and $k_{F}a \simeq
0.15$. As expected in this case, the entire trap can be depleted for
sufficiently long probe duration. Furthermore, the signal decays
exponentially, indicative of a one-body loss process. For the higher
temperature data, the loss is initially exponential, but 25\% of the
initial number remain after a long probe duration.  The remainder
can be understood as the dissociated atom fraction, which is
determined by the fitted temperature $T = 1.8$ $\mu K$ and the
calculated binding energy $E_b = 13$ $\mu$K of the dressed
molecules.  A detailed balance calculation gives approximate
agreement with the observed dissociated fraction. The probe only
weakly couples to free atoms since they have a relatively small
singlet character and excitation occurs only by two-body
photoassociation.

\begin{figure}
\includegraphics[scale=0.35, bb=50 179 712 729]{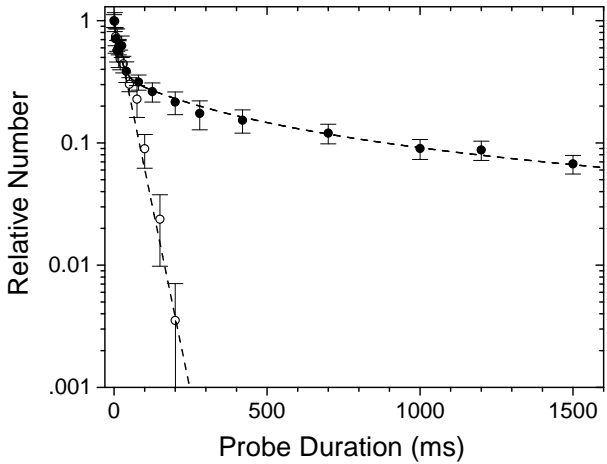}
\caption{\label{loss865}Same as for Fig. \ref{loss695}, except at 865 G.  The
dashed line in the case of the full trap depth data (closed circles)
is a fit to  a ``two-fluid'' model where one component decays via a
rapid one-body loss process and the other via a slower two-body loss
process. Approximately $75\%$ of the gas is lost by the initial fast
process.}
\end{figure}

Figure \ref{loss865} shows the loss of signal vs. probe duration at
 865 G, where $a \simeq -15600~ a_o$. At this field, the gas is in the strongly interacting regime where $k_F|a| > 1$.
According to two-body physics, there are no bound states above the
resonance at 834 G, and probe-induced loss would arise
exclusively from two-body photoassociation. Nonetheless, the
observed loss for the gas prepared in a nearly pure BEC fits to an
exponential rather than to a two-body process. At full trap
depth, the decay consists of two parts: an initial exponential
decay, followed by a much slower two-body process. As the
temperature of the cloud is $T \simeq 0.75$ $T_{F}$, this initial fast
decay may indicate
 the presence of uncondensed paired fermions \cite{chen04} or
  finite-lifetime molecules. The slower (two-body)
process is ascribed to free-bound photoassociation, which is
supported by the fact that the extracted two-body rate coefficient,
$K_2 = 4.9(3.3)\times 10^{-10}$ (cm$^3$ s$^{-1}$)/(W cm$^{-2}$),
agrees well with the calculated value of $9.8(2.6)\times 10^{-10}$
(cm$^3$ s$^{-1}$)/(W cm$^{-2}$) obtained using the expression for
$K_2$ given in Ref.~\cite{Prodan03}, where the uncertainties arise mainly from the temperature determination.

\begin{figure}
\includegraphics[scale=0.35, bb=50 179 712
675]{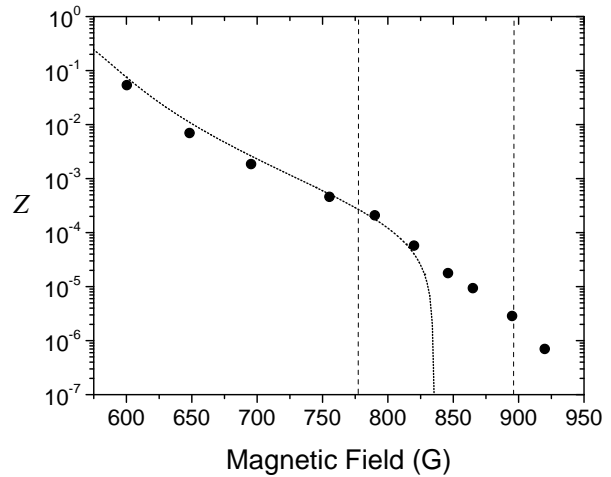} \caption{\label{ZvsB} $Z$ vs. $B$.  The
closed circles represent the value of $Z$ extracted from measured
values of $\Gamma$.  In the case of the $920$ G point, the loss is not
exponential, but the value of $\Gamma$ is taken to be the initial
loss rate. The uncertainty  in $Z$ is
approximately equal to the size of the closed circles, and is due
mainly to uncertainty in the probe laser intensity. The dotted line
shows a comparison with results obtained from a coupled channels
calculation \cite{cch}. The vertical dashed lines represent the
boundaries of the strongly-interacting regime, $k_F|a|
> 1$, where $k_F$ is evaluated using typical values of $N$ at the
low and high field extremes. Although shot-to-shot variations in $N$
are $30\%$, the average value of $N$ at each field is between 13,000
and 90,000 due to day to day variations. $T_F$ is between 200 and
600 nK due to differences in $N$ as well as the trap frequencies.
For all the data, $T<T_c$ and for the points above 850 G, $T<0.5
T_{c}$, where $T_{c}$ refers to the critical temperature at 695 G.
 The gas is expected to be
adiabatically cooled by the increasing field ramp for fields above 755 G
\cite{addiabaticRamp}.}
\end{figure}

Figure \ref{ZvsB} shows the extracted values of $Z$ for fields
between 600 and 920 G. Below 600 G, the dressed molecule lifetime is
too short to obtain a reliable measurement of $Z$. Also shown in
Figure \ref{ZvsB} are the results of a coupled channels calculation,
which represents an exact two-body theory. An analytic
expression for $Z$ on the BEC side of the resonance has been given in Ref.~\cite{falco05} and is in
good agreement with our calculation. While the two-body theory
accurately represents the data for fields below resonance, there is
complete disagreement above resonance. Two-body theory predicts that
$Z$ goes to zero as the resonance is approached, since the size of
the dressed molecules diverges at resonance and produces a vanishing
overlap with the excited molecules. The measured quantity, however,
continues smoothly through resonance, decreasing exponentially with
increasing field. Although the closed-channel fraction is finite and
measurable, its magnitude above resonance is sufficiently small,
$\lesssim$$10^{-5}$, that the expectation of the number of
closed-channel molecules is less than one.

\begin{figure}
\includegraphics[scale=0.35, bb=50 220 712 675]{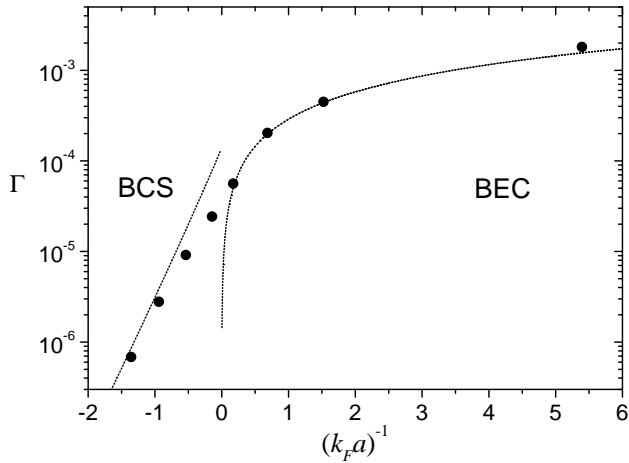}
 \caption{\label{orderparam} Comparison of $\Gamma$ with $|\Delta|^2$. $\Gamma$
 is expressed
 in units of $\Omega^2/\gamma $ making it equivalent to $Z$ plotted in Fig. \ref{ZvsB}.
 The dashed lines correspond to evaluations of $|\Delta|^2$
 in the BCS and the BEC limits integrated over a Thomas-Fermi density
profile.  They have been scaled by the same factor to give the
 best fit on the BEC side. }
\end{figure}

The small closed channel fraction suggests comparison with a single
channel model.  We note that $\Gamma$ is proportional to the local
pair correlation function
$G_{2}(r,r)=\langle\hat{\psi}^{\dagger}_{\downarrow}(r)\hat{\psi}^{\dagger}_{\uparrow}(r)
\hat{\psi}_{\uparrow}(r)\hat{\psi}_{\downarrow}(r) \rangle$, where
$\hat{\psi}_{\uparrow}$ and $\hat{\psi}_{\downarrow}$ are the
fermionic field operators for atoms in different internal states. In
the mean-field approximation $G_{2}$ may be factorized as
$G_{2}(r,r)=n^{2}(r)+
\langle\hat{\psi}^{\dagger}_{\downarrow}(r)\hat{\psi}^{\dagger}_{\uparrow}(r)
\rangle\langle\hat{\psi}_{\uparrow}(r)\hat{\psi}_{\downarrow}(r)
\rangle$, where the first term is the Hartree term with atom density
$n(r)=n_{\uparrow}(r)=n_{\downarrow}(r)$. This term gives rise to a
slow two-body photoassociation process as was observed in the high
temperature data of Fig.~\ref{loss865}. The second term is non-zero
only for correlated pairs and is proportional to $|\Delta|^2$, the
square of the order parameter.
 In the BCS limit, $|\Delta|^2 \propto \epsilon_{F}^{2} e^{-\pi/(k_F
|a|)}$, whereas in the BEC limit, $|\Delta|^2 \propto
\epsilon_{F}^{2}/(k_F a)$ \cite{Engelbrecht97}, which is simply
proportional to $n(r)$, and produces a rapid one-body loss. In Fig.~\ref{orderparam} the data is compared with these functional forms.
The fact that the data have the correct dependence on $(k_F a)^{-1}$
in the BEC and BCS limits is suggestive that such an approach has
captured the essential physics. Data obtained by
starting with higher initial temperatures do not fit the curves
given in Fig.~\ref{orderparam}. Note that non-condensed pairs
\cite{chen04} will give rise to a similar factorization of
$G_{2}$ but it is not expected to have the same dependence on $k_{F}a$.
Although the order parameter presented in Fig.~\ref{orderparam}
comes from a single-channel model, it is the underlying
closed-channel part, albeit small, that gives rise to the detectable
signal, and indeed to the resonance itself.

We have shown that the singlet closed-channel component of the
dressed molecules/pairs is quite small, less than $10^{-3}$,
throughout the strongly interacting regime.  This result strongly
supports the contention of universality for broad resonances.
Contrary to expectations from two-body physics, $Z$ does not vanish
for fields above resonance, but rather is continuous throughout the
resonance regime. Furthermore, the molecular probe distinguishes
between the paired and unpaired components of the gas and has
enabled measurement of the order parameter in BEC and BCS limits, and throughout the strongly interacting regime.

\begin{acknowledgments}
The authors are grateful to Henk Stoof for extensive interaction,
and to Nicolai Nygaard for useful discussions. This work was
funded by grants from the NSF, ONR, NASA, and the Welch Foundation.
\end{acknowledgments}


\begin{thebibliography}{26}
\bibitem{footnote_a_o}{The atomic scattering length $a$ is determined from a coupled channels
calculation \cite{Houb98}, where  we have adjusted our triplet
potential slightly to agree with the recently measured resonance
location of 834 G \cite{Bartenstein05}.}
\bibitem{cch}
{The exact quantity obtained from the coupled channels calculation is \\
$|\langle \psi_{v'=68}(S=0)|\psi_{p}(B)\rangle/\langle
\psi_{v'=68}(S=0)|\psi_{p}(B=0)\rangle|^{2}$.}
\end{thebibliography}


\end{document}